\documentclass[aps,prb,twocolumn,showpacs,superscriptaddress]{revtex4}

\usepackage{graphicx}
\usepackage{amsfonts}
\usepackage{amssymb}
\usepackage{amsmath}
\usepackage{mathrsfs}

\def\v#1{\mathbf{#1}}

\def\kp {\mathbf{k}_\parallel}

\def\ket#1{\vert #1 \rangle}

\begin{document}
\title{Mott scattering at the interface between a metal and a topological insulator}
\author{Erhai Zhao}
\affiliation{Department of Physics and Astronomy, George Mason University, MS 3F3, Fairfax, VA 22030}
\author{Chun Zhang}
\affiliation{
Department of Physics and Department of Chemistry, National University of Singapore, 2 Science Drive 3, Singapore 117542}
\author{Mahmoud Lababidi}
\affiliation{Department of Physics and Astronomy, George Mason University, MS 3F3, Fairfax, VA 22030}

\begin{abstract}
We compute the spin-active scattering matrix and the local spectrum at the interface between a metal and a three-dimensional  topological band insulator. We show that there exists a critical incident angle at which complete (100\%) spin flip reflection occurs and the spin rotation angle jumps by $\pi$. We discuss the origin of this phenomena, and systematically study
the dependence of spin-flip and spin-conserving scattering amplitudes on
the interface transparency and metal Fermi surface parameters.
The interface spectrum contains a well-defined Dirac cone in the tunneling limit, and smoothly evolves into a continuum of metal induced gap states for good contacts. We also investigate the complex band structure of Bi$_2$Se$_3$.
\end{abstract}
\date{\today}
\maketitle

\section{Introduction}
Recently discovered three dimensional topological band insulators \cite{fu07,moore,roy}, such as Bi$_{1-x}$Sb$_x$ \cite{Hsieh2008} and Bi$_2$Se$_3$ \cite{Xia09,zhang2009,Chen09}, are spin-orbit coupled crystal solids with a bulk gap but protected gapless surface states. The low energy excitations at the surface are helical Dirac fermions, i.e., their spin and momentum are entangled (locked) \cite{Hsieh2009}. The charge and spin transport on the surface of a topological insulator
are intrinsically coupled \cite{burkov}.
This makes these materials a promising new platform for spintronics. In addition, heterostructures involving topological insulator, superconductor, and/or ferromagnet have been predicted to show a remarkable array of novel spectral and transport properties (for review
see Ref. \cite{today,rmp,Qi-zhang-rev}). 

Electronic or spintronic devices based on topological insulators will almost inevitably involve metal as measurement probes or functioning components \cite{yokoyama09}. This motivates us to study the local spectrum near the interface between a metal (M) and a topological insulator (TI). For a metal-ordinary semiconductor junction with good contact, it is well known that the metallic Bloch states penetrate into the semiconductor as evanescent waves localized at the interface (for energies within the band gap). Such interface states are known as metal induced gap states (MIGS) \cite{heine65,cohen}. They play an important role in controlling the junction properties, e.g., by pinning the semiconductor Fermi level to determine the Schottky barrier height \cite{tersoff}, a key parameter of the junction.

The local spectrum at the M-TI junction is intimately related to the spin-active scattering of electrons at the M-TI interface. In this paper, we systematically study the evolution of the scattering matrix and the interface spectra
with the junction transparency and metal Fermi surface parameters. 
%by exploiting the {\it complex band structure} of topological insulators, which describes the decaying (rather than propagating Bloch wave) solutions of the crystal Hamiltonian.
The scattering matrix \cite{mrs} we obtain here also forms the basis to investigate the details of the superconducting proximity effect near the superconductor-TI interface \cite{stan}, which was shown by Fu and Kane to host Majorana fermions \cite{majorana}.

The scattering at the M-TI interface differs significantly from its two dimensional analog, the interface between a metal and a quantum spin Hall (QSH) insulator studied by Tokoyama et al \cite{yokoyama09}. They predicted a giant spin rotation angle $\alpha\sim \pi$ and interpreted the enhancement as resonance with the one-dimensional helical edge modes. By contrast, for M-TI interface we predict a critical incident angle at which complete spin flipping occurs and the spin rotation angle jumps by $\pi$. We will explain its origin, {in particular its relation to the surface helical Dirac spectrum}, and discuss its spintronic implications.

This paper is organized as follows. 
We will first compute the scattering matrix using a $\mathbf{k\cdot p}$ continuum model 
by matching the envelope wave functions at the M-TI interface. This simple calculation is easy to understand, 
and it brings out
the main physics of our problem. Along the way, we will discuss the complex band structure of Bi$_2$Se$_3$,
 which describes the decaying (rather than propagating Bloch wave) solutions of the crystal Hamiltonian.
The various caveats of this calculation 
are then remedied by considering a much more general lattice model. Most importantly, it enables us to 
track how the scattering matrix and interface spectrum change with interface transparency. It also sheds light on
the origin of perfect spin-flip scattering at the critical angle.
We will show that the results obtained from these two complementary methods are consistent with each.

\section{Model Hamiltonian and complex band structure}

We consider Bi$_2$Se$_3$ as a prime example of 3D strong topological insulators. Its low energy $\mathbf{k\cdot p}$ Hamiltonian was obtained by Zhang et al \cite{zhang2009},
\[
\hat{H}_{TI}(\v{k})=\epsilon_0(\v{k})\hat{1}+\sum_{\mu=0}^{3}d_\mu(\v{k})\hat{\Gamma}_\mu.
\]
Here $d_0(\v{k})=M-B_1k^2_z-B_2(k_x^2+k_y^2)$, $d_1(\v{k})=A_2 k_x$, $d_2(\v{k})=A_2 k_y$, $d_3(\v{k})=A_1 k_z$, and $\epsilon_0(\v{k})=C+D_1k_z^2+D_2(k_x^2+k_y^2)$. The numerical values of $M$, $A$, $B$, $C$, $D$ are given in Ref. 
\cite{zhang2009}.
We choose the basis ($\ket{+\uparrow}$, $\ket{+\downarrow}$, $\ket{-\uparrow}$,$\ket{-\downarrow}$), where $\pm$ labels the hybridized $p_z$ orbital with even (odd) parity \cite{zhang2009}. The Gamma matrices are defined as
$\hat{\Gamma}_0=\hat{\tau}_3\otimes \hat{1}$, $\hat{\Gamma}_i=\hat{\tau}_1\otimes \hat{\sigma}_i$, with
$\hat{\tau}_i$ ($\hat{\sigma}_i$) being the Pauli matrices in the orbital (spin) space.
The chemical potential of as-grown Bi$_2$Se$_3$ crystal actually lies in the conduction 
band \cite{Hsieh2009}. By hole doping \cite{Hsieh2009} 
or applying a gate voltage \cite{gate}, the chemical potential can be tuned inside 
the gap. The system is well described by $H_{TI}$ (note that energy zero is set as 
in the middle of the band gap).

In this section, we first adopt a rather artificial model for metals with negligible 
spin-orbit coupling. It is
obtained by turning off the spin-orbit interaction (setting $d_\mu=0$ for $\mu$=1,2,3) 
in $H_{TI}$ and shifting the Fermi level into 
the conduction band. The result is spin-degenerate two-band Hamiltonian
\[
\hat{H}_M(\v{k})=[\epsilon_0(\v{k})-E_F]\hat{1}+d_0(\v{k})\hat{\Gamma}_0.
\]
Its band structure, schematically shown in Fig. 1(b), consists of two oppositely dispersing bands 
(the solid and dash line). $E_F$ is tuned to be much higher than the band crossing point, so
the scattering properties of low energy electrons near the Fermi surface are 
insensitive to the band crossing at high energies. This claim will be verified later using a
more generic model for the metal. A similar model was used in the study of metal-QSH interface \cite{yokoyama09}.

%%%%%%%%%%%%%%%%%%%%%%%%%%

\begin{figure}
\includegraphics[width=3.4in]{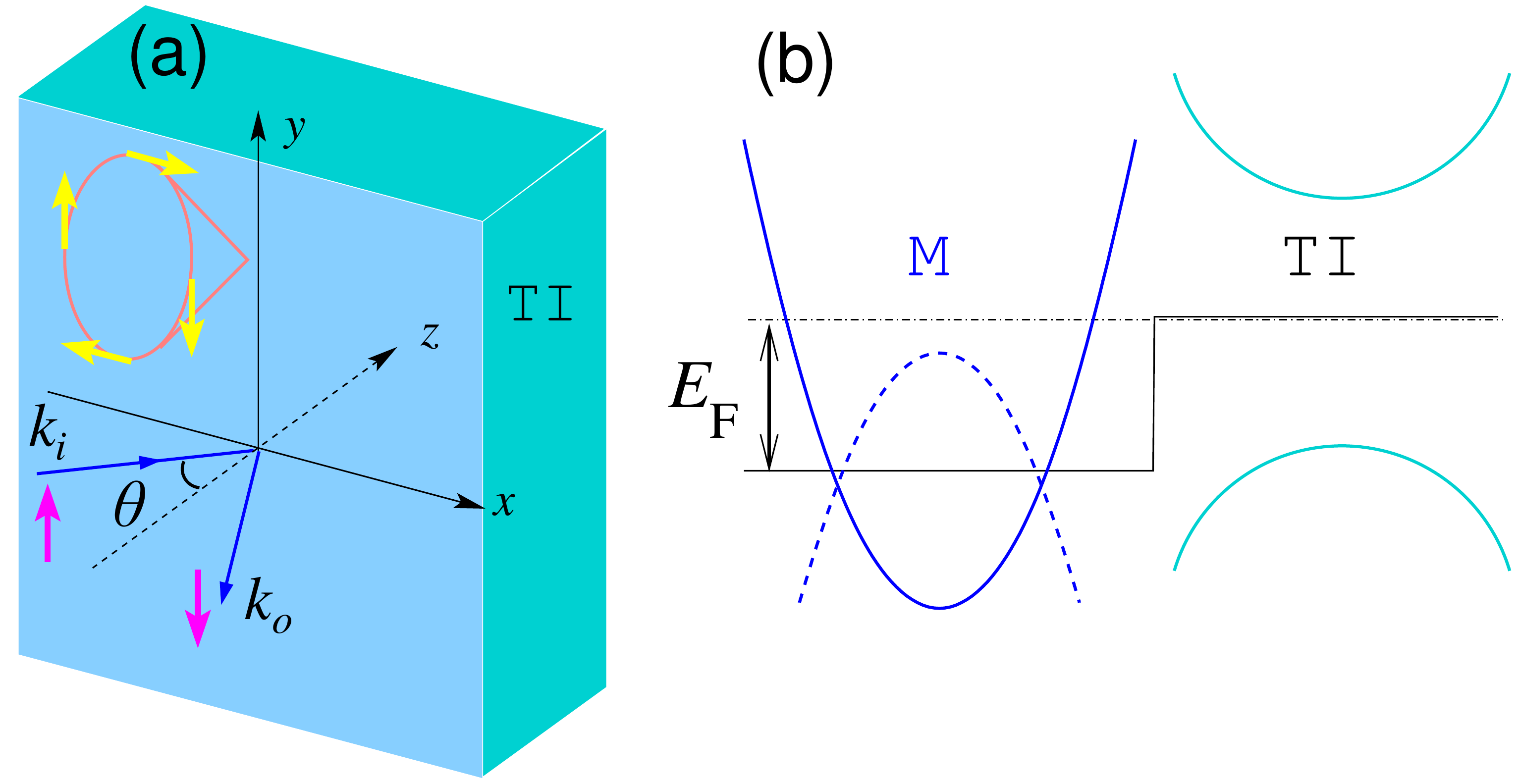}
\caption{(a) Scattering geometry at a metal (M)-topological insulator (TI) interface.
(b) Schematic band structure of the metal (modeled by $\hat{H}_M$) and topological insulator.
}
\end{figure}
%%%%%%%%%%%%%%%%%%%%%%%%%%

Matching the wave functions of two dissimilar 
materials (such as Au and Bi$_2$Se$_3$) at interface is in general 
complicated within the $\mathbf{k\cdot p}$ formalism, because the envelope wave functions 
on either side are defined using different basis (see Ref. \onlinecite{bc} and reference therein). 
For the particular model $H_M$, however, such complication
is circumvented. Then, then wave functions at the metal-TI interface ($z=0$) satisfy the Ben-Daniel 
and Duke boundary condition \cite{duke}, 
\[
\hat{\Phi}_M=\hat{\Phi}_{TI}, \;\;\; \hat{v}_M \hat{\Phi}_M = \hat{v}_{TI}\hat{\Phi}_{TI}.
\]
Here $\hat{\Phi}_i$ is the four-component wave function, and 
the velocity matrix $\hat{v}_{i}=\partial \hat{H}_i/\partial k_z$, $i\in \{M, TI\}$. 
Such boundary condition assumes good atomic contact between two materials.

We are interested in energies below the band gap of TI, so 
$\hat{\Phi}_{TI}$ is evanescent in nature and only penetrates into TI 
for a finite length. Such localized (surface or interface) 
states inside topological insulator can be treated within the $\mathbf{k\cdot p}$ formalism 
using the theory of {\it complex band structures}, pioneered by Kohn \cite{kohn59}, Blount \cite{blount62}, 
and Heine \cite{heine63} et al. 
The main idea is to allow the crystal momentum to be complex and analytically continue
$H_{TI}(\v{k})$ to the complex $\v{k}$ plane. 
While the extended Bloch waves are the eigen states of $H_{TI}(\v{k})$ for real $\v{k}$, 
eigen functions of $H_{TI}(\v{k})$ for complex $\v{k}$ describe localized states. Together they
form a complete basis to describe crystals of finite dimension. 

%%%%%%%%%%%%%%%%%%%%%%%%%%
\begin{figure}
\includegraphics[width=1.7in]{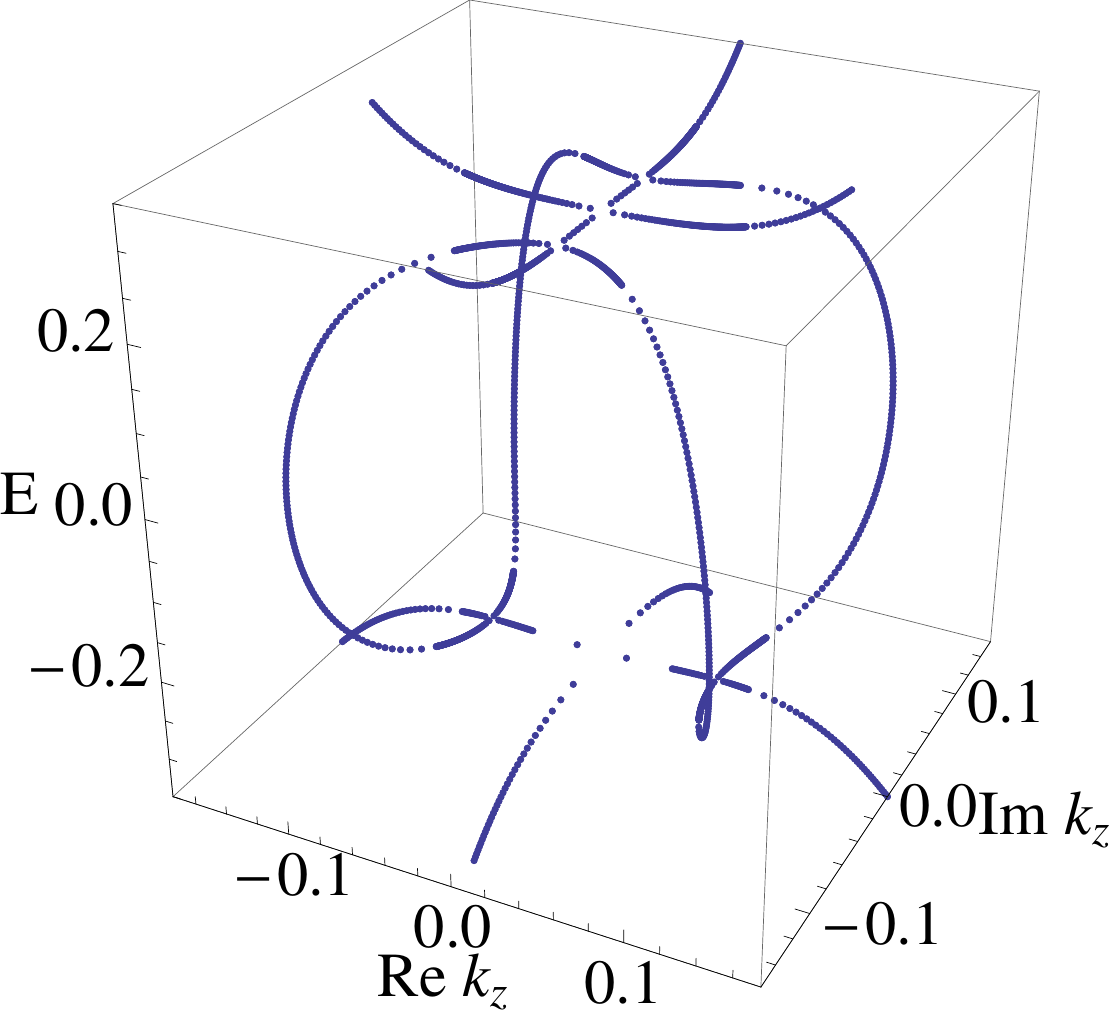}\includegraphics[width=1.7in]{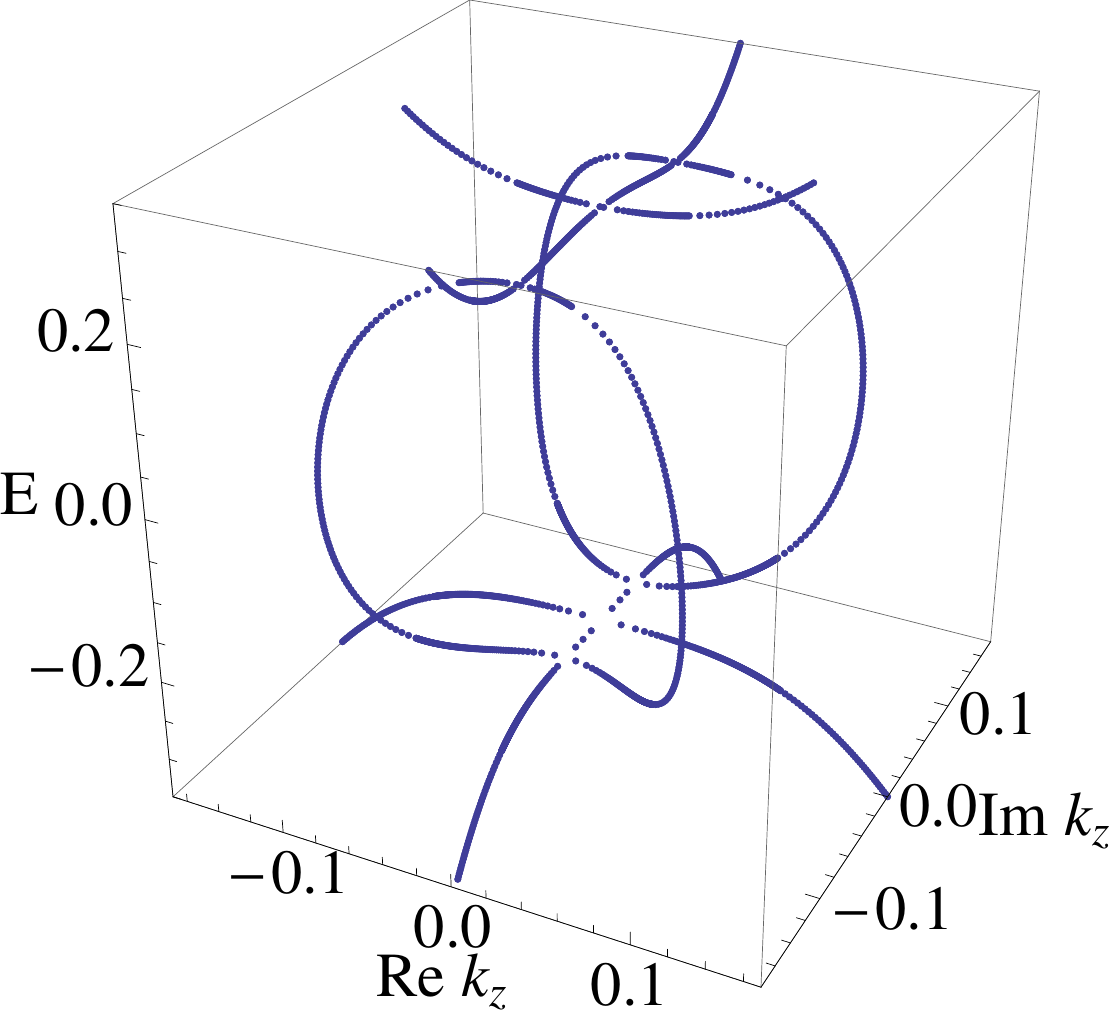}
\caption{The complex band structure
of topological insulator described by $\hat{H}_{TI}(\v{k})$ 
for $k_y=0$, $k_x=0.02$ (left) and $0.04$ (right). $E$ is measured in eV, and $k$ in $\AA^{-1}$.
Subgap states with complex $k_z$ represent evanescent waves. 
The topology of  real lines \cite{heine63}  changes as $k_x$ is increased.  
}
\end{figure}
%%%%%%%%%%%%%%%%%%%%%%%%%%

In our scattering problem, we have to find all eigen states of $H_{TI}(\v{k})$ with energy $E$ and 
wave vector $\v{k}=(k_x,k_y,\tilde{k}_z)$, where $k_x$ and $k_y$ are given and real, but $\tilde{k}_z$ is 
complex and unknown. For a general $\mathbf{k\cdot p}$ Hamiltonian such as $\hat{H}_{TI}$, 
we follow Chang and Schulman \cite{chang82} to rewrite it as
\[
\hat{H}_{TI}=\hat{h}_0(k_x,k_y)+\hat{h}_1 \tilde{k}_z+\hat{h}_2\tilde{k}^2_z,
\]
where $\hat{h}_1=A_1\hat{\Gamma}_3$, and $\hat{h}_2=-B_1\hat{\Gamma}_0$. 
Then the eigen equation $(\hat{H}_{TI}-E\hat{1})\hat{\phi}=0$ can be reorganized into an 
eigen value problem for $\tilde{k}_z$,
\[
\left(
\begin{array}{ll}
  0 & 1   \\
  -\hat{h}_2^{-1}(\hat{h}_0-E\hat{1}) & -\hat{h}_2^{-1}\hat{h}_1
  \end{array}
\right)
\left(
\begin{array}{l}
  \hat{\phi}   \\
  \hat{\phi}'  
\end{array}
\right)
=\tilde{k}_z \left(
\begin{array}{l}
  \hat{\phi}   \\
  \hat{\phi}'    
\end{array}
\right).
\]
%%% note: these details can be omitted to save space 
Then all possible values of $\tilde{k}_z$ can be obtained for given incident parameter $E$, $k_x$, and $k_y$. 
For the anisotropic Dirac Hamiltonian $H_{TI}(\v{k})$, the energy eigenvalues can be obtained 
analytically \cite{qi_field}, which allows for an analytical solution of the complex band structure.

For $E$ within the gap, there are in general 4 pairs of complex solution of $\tilde{k}_z$, for if $\tilde{k}_z$ is a solution so is $\tilde{k}^*_z$. 
We label those with positive imaginary parts with $\{\tilde{k}^\nu_z\}$, and the corresponding wave function $\{\hat{\phi}^\nu \}$, $\nu=1,2,3,4$. They are decaying solutions in the half space $z>0$. In our model, $\tilde{k}_z$ turns out to be doubly degenerate, as shown in Fig. 2. The wave function inside TI ($z>0$) then has the form
\[
\hat{\Phi}_{TI}=\sum_{\nu} t_\nu e^{i\tilde{k}^\nu_z z} \hat{\phi}_\nu.
\]

\section{Scattering matrix from wave-function matching} 

To set the stage for discussing scattering off a topological insulator, it is instructive to recall the generic features of elastic scattering of electrons by a heavy ion with spin-orbit interaction. This classical problem was solved by Mott, and known as {\it Mott scattering}. 
%It is being used, for example, to measure the spin polarization of photo electrons in spin-resolved ARPES. 
% Spin-orbital coupling amounts to a momentum dependent magnetic field $\mathbf{B}(\mathbf{k})$, which causes the spin of incident electron to precess. 
The scattering matrix has the general form \cite{mott}
\[
\hat{S}_{Mott}=u\hat{1}+w\hat{\boldsymbol{\sigma}}\cdot (\mathbf{k}_{i}\times \mathbf{k}_{o}),
\]
where $\mathbf{k}_{i}$ and $\mathbf{k}_{o}$ are the incident and outgoing momentum respectively, $\hat{\boldsymbol{\sigma}}$ is the Pauli matrix, and $u,w$ depend on the scattering angle. It is customary to
define the spin-flip amplitude $f=S_{21}$, and spin-conserving amplitude $g=S_{11}$. 
Both $f$ and $g$ are complex numbers, their relative phase defines the {\it spin rotation angle} $\alpha=\mathrm{Arg}(g^*f)$.
One immediately sees that for back scattering, $\hat{S}_{Mott}=u\hat{1}$, so there is no spin flip, $f=0$. As we will show below, this also holds true for scattering off TI.

Now consider an electron coming from the metal 
with momentum $\v{k}$ incident on the M-TI interface located at $z=0$, 
as schematically shown in Fig. 1(a). 
We assume the interface is 
translationally invariant, so the transverse momentum $\v{k}_{\parallel}=(k_x,k_y)$ is 
conserved, and the energy $E$ of the electron 
lies within the band gap of TI. Then, only total reflection 
is possible, but the spin-orbit coupling inside TI acting like a $\v{k}$-dependent magnetic field rotates the spin of the incident particle. The scattering (reflection) matrix has the form
\[
\hat{S}(\v{k})=\left(
\begin{array}{ll}
  g & \bar{f}   \\
  f & \bar{g}
  \end{array}
\right),
\]
where $|g|^2+|f|^2=1$. 
Our goal is to find the dependence of the scattering amplitudes $f,g$ 
on $\v{k}$, or equivalently, on energy $E$ and 
incident angle $\theta$. From time-reversal symmetry, 
$\bar{f}(E,\theta)=f(E,-\theta)$ and $\bar{g}(E,\theta)=g(E,-\theta)$.
We shall show that
$f(\v{k}_{\parallel})=-f(-\v{k}_{\parallel}$),
$g(\v{k}_{\parallel})=g(-\v{k}_{\parallel}$). So 
$f$ is an odd function of $\theta$, while $g$ is even in $\theta$.  
Since our problem can be viewed as coherent multiple scattering from a lattice 
array of Mott scatters occupying half the space, we will refer to
spin-active scattering at the metal-TI interface as Mott scattering.

Consider a spin up electron from the conduction band of the metal 
with momentum $\v{k}$ and energy $E=
\epsilon_0(\v{k})-E_F-d_0(\v{k})$ lying within the band gap of TI.
The wave function inside the metal ($z<0$) has the form
\[
\hat{\Phi}_M=(r_1e^{-ik'_{z} z},r_2e^{-ik'_{z}z},e^{ik_{z}z}+r_3 e^{-ik_{z}z} ,r_4 e^{-ik_{z}z})^{\mathrm{T}},
\]
up to the trivial $e^{i(k_x x+k_y y)}$ and renormalization factor.
Here $k_z=\hat{z}\cdot\v{k}$, and $\{r_i\}$ are the reflection amplitudes. We identify 
the spin flip amplitude $f=r_4$ and the spin-conserving amplitude $g=r_3$. Note that 
there is no propagating mode at energy $E$ available in the valence band 
for the reflected electron. So $k'_z$ is purely imaginary. 
At such energy, there is no propagating mode available in TI. We have discussed the 
evanescent wave function $\hat{\Phi}_{TI}$ in the previous section.
With $\hat{\Phi}_{M}$ and $\hat{\Phi}_{TI}$, we solve the boundary condition at $z=0$ 
to obtain $r_\nu, t_\nu$ and the scattering matrix $S$. 

Fig. 3 shows the magnitude and phase of $f$ and $g$ versus the incident angle $\theta$ for $E=0.1$eV, with $E_F$ set to be 0.28eV. At normal incidence, $\theta=0$, spin flip scattering
is forbidden as in the single-ion Mott scattering. With increasing $\theta$ the magnitude of $g$ drops continuously. At a critical angle $\theta_c$, $|g|$ drops to zero and we have perfect (100\%) spin flip reflection.
At the same time, the spin rotation angle $\alpha$ (the relative phase between $f$ and $g$)
jumps by $\pi$.

It is tantalizing to think of what happens at $\theta_c$ as resonant scattering
with the helical surface mode of the TI. This however is problematic.
We are considering good contacts at which the wave functions of the two materials hybridize strongly. 
Surface mode is preempted by MIGS. Indeed, we checked that the corresponding critical 
transverse momentum $k_\parallel$ depends only weakly on $E$. This is at odds with
the linear dispersion of the TI surface mode, $E=A_2k_\parallel$ \cite{zhang2009}. To gain better
understanding, we now switch to a lattice model to systematically study the role of interface 
transparency and metal Fermi surface parameter ($E_f, k_f, v_f$) on the scattering matrix.

%%%%%%%%%%%%%%%%%%%%%%%%%%
\begin{figure}
\includegraphics[width=2.5in]{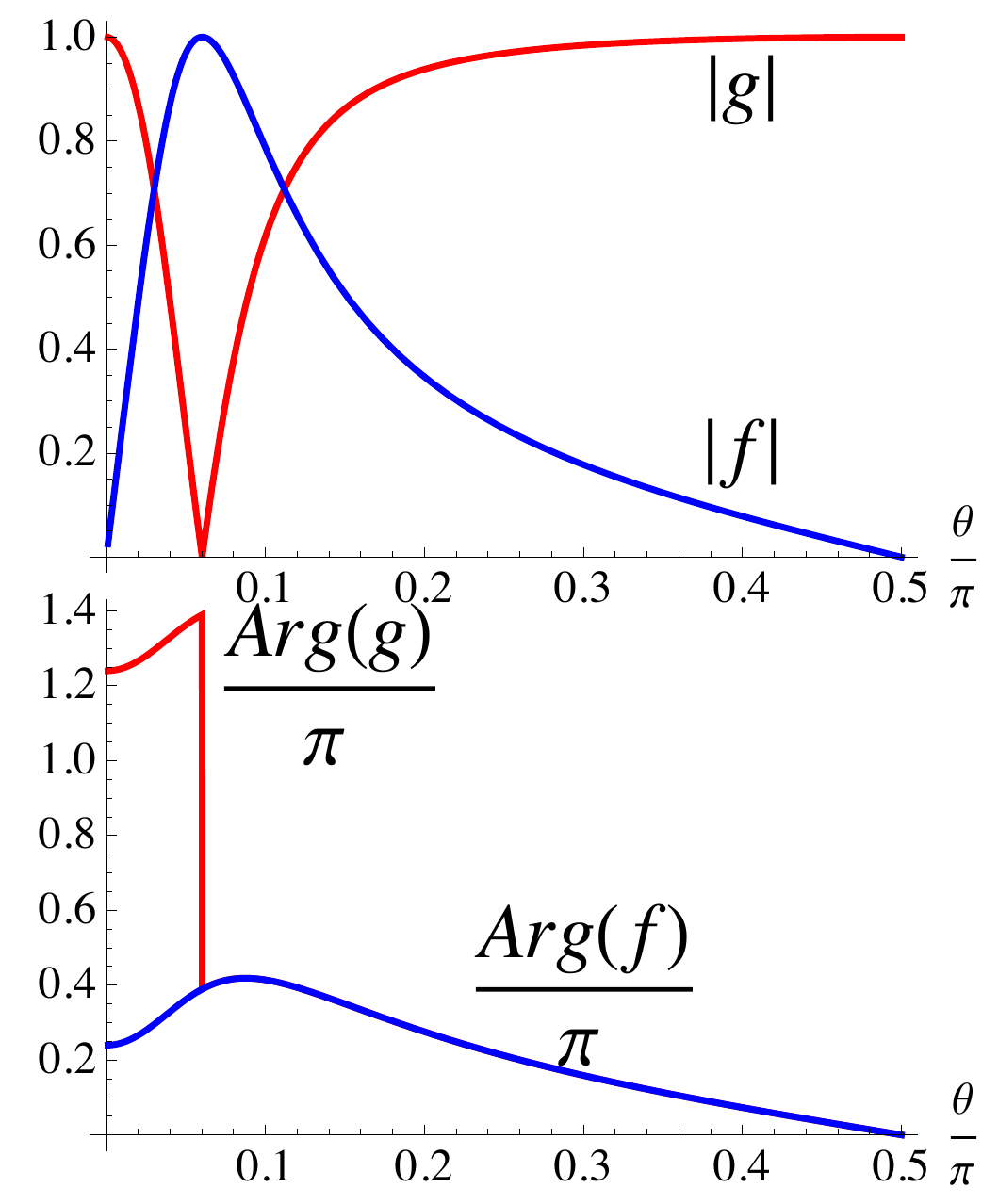}
\caption{ The magnitudes (upper panel) and the phases (lower panel) of the spin-flip 
amplitude $f$ and spin-conserving amplitude $g$ versus the incident angle $\theta$.
$E=0.1$eV, $E_F$=0.28eV. $|g|^2+|f|^2=1$. Arg($g$) and Arg($f$) 
are shifted upward by $\pi$ for clarity.}
\end{figure}
%%%%%%%%%%%%%%%%%%%%%%%%%%

\section{Interface spectrum and scattering matrix from lattice Green function} 

We consider a simple lattice model for the M-TI junction.
The topological insulator is modeled by 
a tight binding Hamiltonian on cubic lattice,
\begin{align*}
&\mathscr{H}_R=\sum_{\kp,n}\left\{ 
\hat{\psi}_{\kp,n}^\dagger (b_1\hat{\Gamma}_0-i\frac{a_1}{2}\hat{\Gamma}_3)  \hat{\psi}_{\kp,n+1}+ h.c. \right. \\
&+\left. 
\hat{\psi}_{\kp,n}^\dagger\left[d(\kp)\hat{\Gamma}_0+a_2(\hat{\Gamma}_1\sin k_x +\hat{\Gamma}_2\sin k_y)\right] \hat{\psi}_{\kp,n} \right\} .
\end{align*}
Here $\hat{\psi}=(\psi_{+\uparrow},\psi_{+\downarrow},\psi_{-\uparrow},\psi_{-\downarrow})^\mathrm{T}$ is the annihilation operator, $d(\kp)=M-2b_1+2b_2(\cos k_x+\cos k_y-2)$ with $k$ measured in $1/a$. 
The cubic lattice consists of layers of square lattice stacked in the $z$ direction,
$n$ is the layer index, and $\kp$ is the momentum in the $xy$ plane.
The isotropic version of $\mathscr{H}_R$, with $a_1=a_2$, $b_1=b_2$, was 
studied by Qi et al as a minimal model for 3D topological insulators \cite{qi_field}.
To mimic Bi$_2$Se$_3$, we set the lattice spacing $a=5.2$\AA, which gives the correct unit cell volume, 
and $a_i=A_i/a$, $b_i=B_i/a^2$ for $i=1,2$. Although a crude caricature 
of the real material, $\mathscr{H}_R$ yields the correct gap size and surface dispersion, it also reduces to 
the continuum  $\mathbf{k\cdot p}$ Hamiltonian $\hat{H}_{TI}$ in the small $k$ limit, 
aside from the topologically trivial $\epsilon_0(\v{k})$ term.
 
%%%%%%%%%%%%%%%%%%%%%%%%%%
\begin{figure}
\includegraphics[width=1.7in]{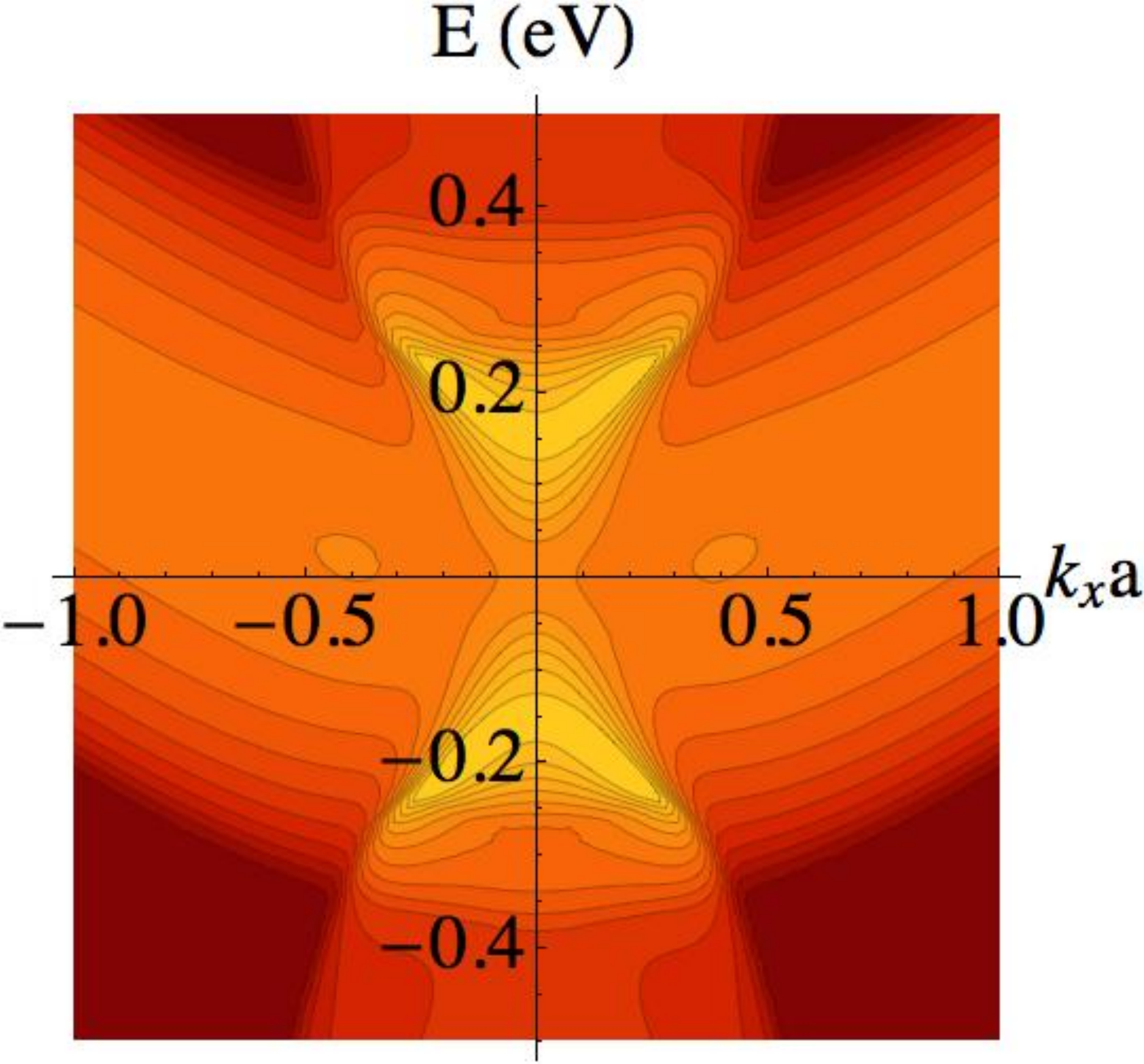}\includegraphics[width=1.7in]{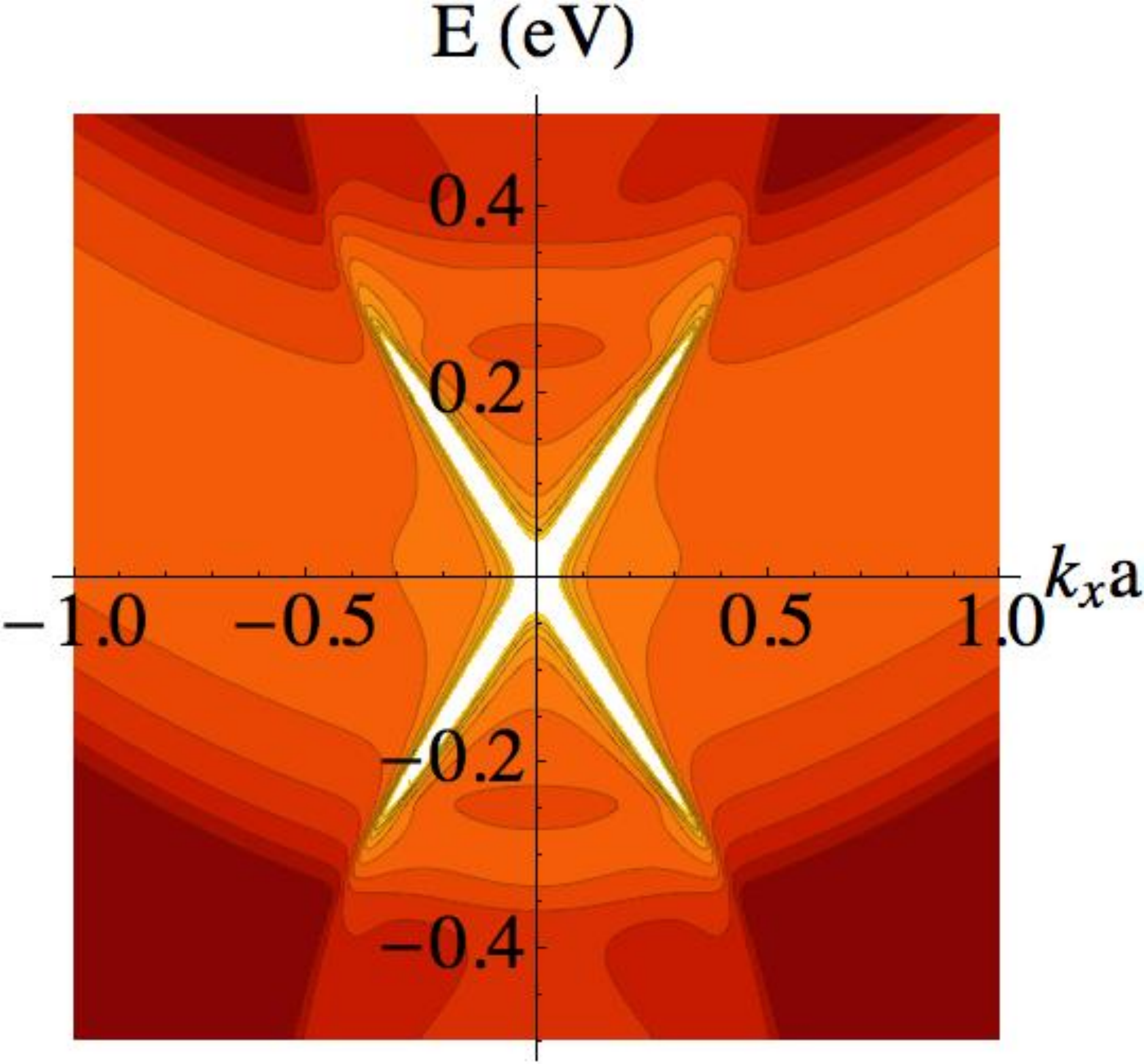}
\caption{The spectral function $N(E,k_x,k_y=0)$ at the interface of metal and topological insulator. Left: 
good contact, $J=t_M$, showing the continuum of metal induced gap states. 
Right: poor contact with low transparency, $J=0.2t_M$, showing well 
defined Dirac spectrum as on the TI surface. $t_M=0.18eV$, $\mu_M=-4t_M$, $a$ is lattice spacing.}
\end{figure}
%%%%%%%%%%%%%%%%%%%%%%%%%%
 
As a generic model for metal, we consider a single band tight binding Hamiltonian on cubic lattice,
\[
\mathscr{H}_L=\sum_{\kp,n,\sigma}[h({\kp})  n_{\kp,n,\sigma}
- t_M \phi_{\kp,n,\sigma}^\dagger  \phi_{\kp,n+1,\sigma} + h.c.] 
\]
where $h(\kp)=-2t_M(\cos k_x+\cos k_y)-\mu_M$. 
The Fermi surface parameters of the metal can be varied by tuning $t_M$ and $\mu_M$.
The metal occupies the left half space, $n\leq 0$, and 
the TI occupies the right half space $n\geq 1$. The interface domain consists of layer $n=0,1$. 
The coupling between metal and TI is described by hopping,
\[
\mathscr{H}_{LR}=-\sum_{\kp,\ell,\sigma}J_{\ell}\psi^{\dagger}_{\kp,n=1,\ell,\sigma}\phi_{\kp,n=0,\sigma}+h.c.
\]
$J_{\ell}$ is the overlap integral between the $p$-orbital $\ell=\pm$ of TI and the $s$-like orbital of metal. For simplicity, we assume $J_{\ell}$ is independent of spin. Then, $J_{+}=-J_{-}=J$. $J$ can be tuned from weak to strong. Small $J$ mimics a large tunneling barrier between M and TI, 
and large $J$ (comparable to $t_M$ or $B_2$) describes a good contact. 

%%%%%%%%%%%%%%%%%%%%%%%%%%
\begin{figure}
\includegraphics[width=1.7in]{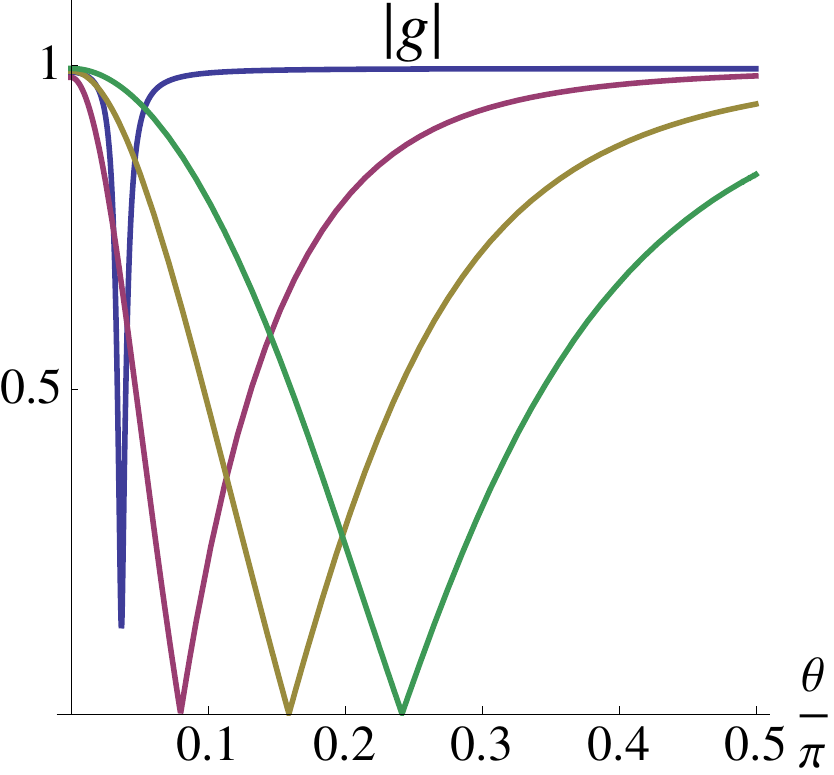}\includegraphics[width=1.7in]{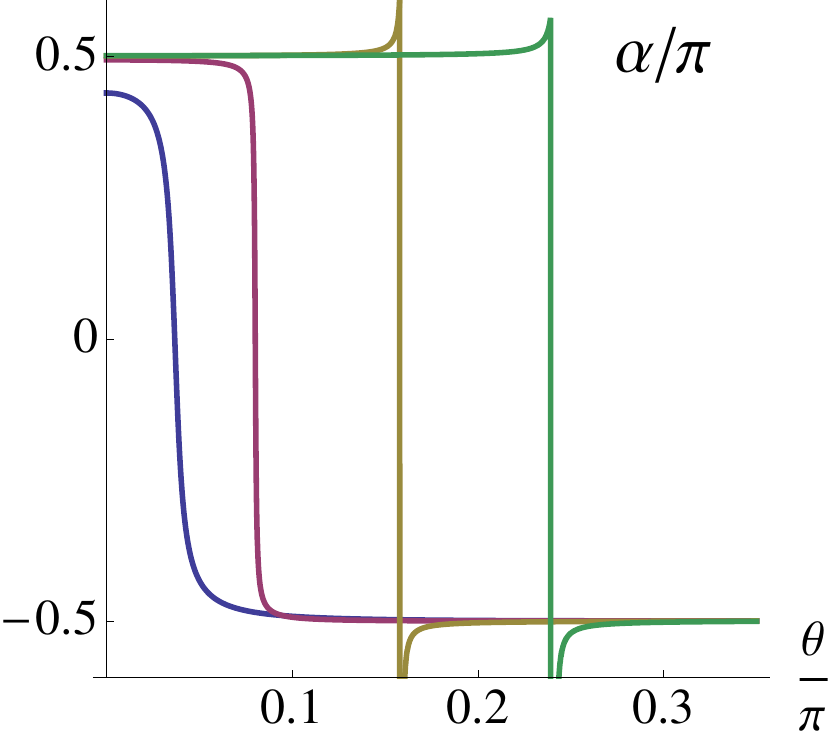}
\caption{The spin-conserving reflection amplitude $|g|$ and spin rotation angle 
$\alpha$ versus the incident angle $\theta$ for increasing contact transparency,
$J/t_M=0.25, 1, 1.5, 2$ (from left to right). $t_M=0.18eV$, $\mu_M=-4t_M$, $E=0.05$eV, $k_y=0$.
$|f|^2=1-|g|^2$. 
}
\end{figure}
%%%%%%%%%%%%%%%%%%%%%%%%%%

The lattice Green function of the composite system is computed via 
standard procedure by introducing the inter-layer transfer matrix 
and the method of interface Green function matching \cite{gf}. 
Fig. 4 shows two examples of the local spectral function 
(momentum-resolved density of states) at the interface,
\[ 
N(E,\kp)=-\sum_{n=0,1}\mathrm{Im Tr}\hat{\mathscr{G}}(E,\kp)_{n,n}, 
\]
where $\hat{\mathscr{G}}(E,\kp)_{n,n'}$ is 
the local Green function at the interface with $n,n'=0,1$, and  
the trace is over the spin and orbital space. 
In the tunneling (weak coupling, small $J$) limit, the interface spectrum includes 
a sharply defined Dirac cone as on the surface of TI. As $J$ is increased, 
the linearly dispersing mode becomes ill defined and eventually replaced 
by a continuum of metal induced gap states.

Once the lattice Green function is known for given incident $E$ and $k_{\parallel}$, 
the scattering (reflection) matrix can be constructed from $\hat{\mathscr{G}}$ by \cite{gf},
\[
\hat{S}(E,\kp)=\hat{\mathscr{G}}(E,\kp)_{0,0}g_M^{-1}(E,\kp)-\hat{1}
\]
where $g_M$ is the spin-degenerate bulk Green function of metal. Fig. 5 shows
the evolution of $|g(\theta)|$ and $\alpha(\theta)$ for increasing $J$, where 
a level broadening of $E/10$ is used. Most importantly,
we observe that the existence of a critical angel $\theta_c$, 
where complete spin-flip occurs and $\alpha$ jumps by $\pi$, is a robust phenomenon. 
It is independent of the details of the contact, the metal Fermi 
surface, or other high energy features in the band structure.

To understand the perfect spin flip, we first focus on 
the tunneling limit, $J\ll t_M$. In this limit, the local spectrum at layer $n=1$ as shown
in the right panel of Fig. 4 approaches
the TI surface spectrum, namely the helical Dirac cone. An incident up spin 
tunneling across the barrier will develop resonance with the helical mode, which is
a quasi-stationary state with long life time, if its 
momentum and energy satisfy $k_\parallel=E/A_2$. Moreover, it has to flip its spin, since 
only down spin can propagate in the $k_x$ direction (suppose $k_y=0$). 
The $\pi$ jump in the phase shift is also characteristic of the resonance. Indeed,
we have checked that precisely at $\theta_c$ the resonance criterion, $k_f\sin\theta_c=E/A_2$, 
is met.
We also varied $\mu_M$ for fixed $J$ and $t_M$, bigger $\mu_M$ 
yields a bigger Fermi surface and a smaller $\theta_c$. This is consistent with 
the resonance criterion above.

As $J$ is increased, the width of the resonance grows and eventually it is replaced by a 
broad peak (dip) in $|f|$ ($|g|$), but the vanishing of $|g|$ and $\pi$
shift in $\alpha$ at $\theta_c$ persist to good contacts,
even though in this limit the interface is flooded by MIGS (left panel of Fig. 4) 
and bears little resemblance to the Dirac spectrum.
With all other parameters held fixed, $\theta_c$ increases with $J$. Qualitatively,
coupling to TI renormalizes the metal spectrum near the interface, producing a smaller effective $k_f$ (hence a larger $\theta_c$) compared to its bulk value. 
It is remarkable that perfect spin flip at the critical angle persists all
the way from poor to good contacts. Indeed, the main features observed here for
for good contacts using the lattice model 
agree well with the results obtained in previous section by wave function matching. 

\section{discussions}

We now discuss the experimental implications of our results. The M-TI interface spectrum can be measured by 
ARPES (or scanning tunneling microscope) experiments on metal film coated on a topological insulator.
Our results also suggest that a topological insulator can serve as a perfect mirror to flip the 
electron spin in metal. Such spin-active scattering at the M-TI interface may be 
exploited to make novel spintronic devices. The magnitude of $g$ or $f$
can be measured by attaching two ferromagnetic leads to a piece of metal in contact with TI, 
forming a multi-terminal device. 
One of the ferromagnetic leads produces spin-polarized electrons incident on the M-TI interface at some angle, while 
the other lead detects the polarization of reflected electron, as in a
giant magneto-resistance junction. The spin rotation angle $\alpha$ can be measured 
indirectly by comparing the predicted current-voltage characteristics of M-TI-M 
or Superconducto-TI-Superconductor junctions, which are sensitive the phase shift $\alpha$. It can also be inferred from 
the spin transport in a TI-M-TI sandwich, as discussed for QSH insulator in Ref. \cite{yokoyama09}.
Detailed calculations of the transport properties of these structured,
using the scattering matrix obtained here, will be subjects of future work.

\section{acknowledgements}
We thank Liang Fu, Parag Ghosh, Predrag Nikolic, Indu Satija, and Kai Sun 
for helpful discussions. This work is supported by NIST Grant No. 70NANB7H6138 Am 001 
and ONR Grant No. N00014-09-1-1025A (EZ). 

\bibliographystyle{apsrev}
\bibliography{topological}

\end{document}